\documentclass{llncs}

\usepackage[super]{nth}
\usepackage{amssymb}
\usepackage{amsmath}
\usepackage{graphicx}
\usepackage[american]{babel}
\usepackage{url}

\newcommand{\vex}[1]{\boldsymbol{#1}}

\usepackage[babel]{csquotes}
\usepackage{booktabs}

\providecommand{\acknowledgements}[1]{\vskip1.5ex\noindent \textbf{Acknowledgements.} #1}

\title{Basis-Function Modeling of Loudness Variations in Ensemble Performance}

\author{Thassilo Gadermaier, Maarten Grachten, and Carlos Eduardo Cancino Chac\'on}
\institute{Austrian Research Institute for Artificial Intelligence\\\email{firstname.lastname@ofai.at}}

\begin{document}

\maketitle

\begin{abstract}This paper describes a computational model of loudness variations in expressive ensemble performance.
The model predicts and explains the continuous variation of loudness as a function of information extracted automatically from the written score.
Although such models have been proposed for expressive performance in solo instruments, this is (to the best of our knowledge) the first attempt to define a model for expressive performance in ensembles.
To that end, we extend an existing model that was designed to model expressive piano performances, and describe the additional steps necessary for the model to deal with scores of arbitrary instrumentation, including orchestral scores.
We test both linear and non-linear variants of the extended model on a data set of audio recordings of symphonic music, in a leave-one-out setting.
The experiments reveal that the most successful model variant is a recurrent, non-linear model.
Even if the accuracy of the predicted loudness varies from one recording to another, in several cases the model explains well over 50\% of the variance in loudness.
\end{abstract}


\section{Introduction}\label{sec:introduction}
This paper describes a computational model of loudness variations in ensemble performance.
We are primarily interested in the expressive factors that influence loudness.
Although expression is a very broad term that may include the mental or physical state of the performer(s), their communicative intentions, the targeted audience, and so on, we focus on those aspects of expression that are determined by the written score.
In other words, the proposed model is intended to account for the ways in which information extracted from the written score influences the continuous variation of loudness throughout the recording of a performance.

The potential uses of such a model are twofold.
Firstly, its predictive capacities can be used to generate more natural, musically appropriate acoustic renderings of a piece, than a straight-forward mechanical rendering.
Such improved renderings can also improve tasks such as offline score-performance alignment, and \emph{automatic live score-following}---a scenario in which a computer keeps track of musical time during the performance of a piece of music~\cite{dblp:dblp_conf/ecai/arztwd08}.

Secondly, a model of expressive loudness variations may also be used for explanatory purposes.
This means that the model can attribute variations in the expressive quality of a performance to factors like performance directives that were written in the score by the composer (like \emph{crescendo}, \emph{diminuendo}, and \emph{fermata}), and other aspects of the written score.
Explanatory visualizations of expressive performances based on this information can be used for didactic purposes, to introduce an audience to the phenomenon of expressive music interpretation.

As the point of departure for the model of expressive ensemble performance proposed here, we take an existing framework, the \emph{basis-function} modeling approach, which has been successfully used in modeling solo piano~\cite{grachten12:_linear}.
In addition to this linear version of the model, improved results have been obtained using non-linear variants. The non-linear variants can model more complex relationships between expressive parameters and the score, as demonstrated in \cite{chacon15}, where the basis-function representation is combined with a \emph{feed-forward neural network} (FFNN). 
A more sophisticated form of non-linear modeling involves \emph{recurrent} network connections, allowing for temporal dependencies in the relation between score information and expressive parameters. This type of model was shown to outperform non-temporal models for predicting expressive timing in classical piano performances~\cite{grachten16:brnbm}.
In the current paper, we employ both the linear, and the two non-linear variants of the model.

The main contribution of the current paper is the extension of the basis-function definition to accommodate for ensembles of instruments, possibly including multiple instances of the same instrument, as is common in orchestral scores.
We discuss the difficulties and complications that arise when dealing with recordings of large ensembles, rather than a single piano.
To address these issues, we define \emph{merging} and \emph{fusion} operations on basis-function representations, as explained further on in the paper.
These operations are needed to train the model on pieces with different instrumentations, and to present the score information of the joint orchestral score to the model in a unified way.

We evaluate the proposed basis-function model for ensemble performance using a dataset of 16 orchestral recordings of pieces by Bruckner, Mahler, and Beethoven, as performed by the Royal Concertgebouw Orchestra.
The results show that depending on the piece, a considerable part of the total variance in loudness can be explained by information from the score. Furthermore, there are notable differences between the models, with the non-linear models, especially the recurrent model, performing better than the simpler linear model.

In Section~\ref{sec:related-work-state}, we give a very brief overview music expression research, focusing on computational approaches. Section~\ref{sec:comp-expr-model} covers the description of the proposed model for ensemble performance. An experimental validation of the model is described in Section~\ref{sec:experiments}, including the presentation and discussion of results. Finally, conclusions are formulated in Section~\ref{sec:conclusion}.

\section{Related work and state of the art}\label{sec:related-work-state}
Empirical research and modeling of musical expression have a long history, with accounts of measurements of music performances dating back as far as the late \nth{19} century~\cite{binet96}, and the first half of the \nth{20} century~\cite{seashore38}.
Despite these early precursors, expressive performance research has gained substantial traction only since the 1980s, presumably incited in part by the advent of modern computers, electronic instruments, and the corresponding MIDI protocol for transmission of musical information, facilitating the recording of performances, and subsequent analysis of the data obtained in this way.

A significant number of empirical studies have sought to establish relationships between some aspects of expression and particular explanatory factors.
These factors can be roughly divided into those that relate to the performer's intention of expressing particular emotions, and those that relate to the musical structure, in the broadest sense of the word.
For example, a widely confirmed mapping between emotion and expression is that slow tempo, legato articulation, and softer timbres contribute to the perception of the music as sad or solemn, whereas a faster tempo, staccato articulation and brighter timbres tend to induce a perception of happiness \cite{juslin97:_emotion_commun_music_perfor}, \cite{juslin01a}, \cite{canazza03a}.
Similarly, various structural aspects of the musical score have been found to influence musical expression~\cite{clarke91}.
Most notably, musical grouping structure (the division of the music into \emph{motifs}, and \emph{phrases}) is often expressed in arc like shapes in tempo and dynamics \cite{todd89}.
Another type of musical structure that musicians express through expressive variations is the metrical structure~\cite{sloboda83}.

Research on expression in ensemble performance is sparse.
Studies in this area often focus on synchronization between musicians \cite{keller14:_ensem}, \cite{goebl2009synchronization}, and the cues musicians use to communicate and synchronize \cite{Bishop:2014ve}, \cite{glowinski2014he}.

\subsection{Computational modeling of musical expression}\label{sec:comp-models-music}
Computational models of expressive music performance seek to clarify the relationships between certain properties of the musical score and performance context with the actual performance of the score~\cite{widmer04:_comput_model_expres_music_perfor}.
These  models can serve mainly analytical purposes~\cite{widmer02,windsor97}, by showing the relation between structural properties of the music and its effect in the performance of such music, mainly predictive purposes~\cite{teramura08:_gauss_proces_regres_render_music_perfor}, i.e.\ the models are used to render expressive performances, or both~\cite{grindlay06}, \cite{depoli01,grachten12:_linear}.
Computational models of music performance tend to follow two basic paradigms: \emph{rule based} approaches, where the models are defined through music-theoretically informed rules that intend to map structural aspects of a music score to quantitative parameters that describe the performance of a musical piece,   and \emph{data-driven} (or \emph{machine learning}) approaches, where the models try to infer the rules of performance from analyzing patterns obtained from (large) datasets of observed (expert) performances~\cite{widmer03}.

One of the most well-known rule-based systems for musical music performance was developed at the Royal Institute of Technology in Stockholm (referred to as the KTH model)~\cite{friberg06:_overv_kth_rule_system_music_perfor}.
This system is top-down approach that describes expressive performances using a set of (music theoretically sound/cognitively plausible) performance rules  that predict aspects of timing, dynamics and articulation, based on a local musical context.

Among the machine learning methods for musical expression is the model proposed by \cite{bresin98}.
This model  uses artificial neural networks (NNs) in a supervised fashion in two different contexts: 1) to learn and predict the rules proposed by the KTH model and 2) to learn the performing style of a professional pianist using an encoding of the KTH rules as inputs.
Similarly, the \emph{basis-function modeling approach} (see Section {\ref{sec:basis-functions}}) used by \cite{grachten12:_linear} and \cite{chacon15} represents a bottom-up  approach that uses a lower level encoding of a musical score in order to learn how different aspects of the score contribute to generate an expressive performance of a musical piece.

Grachten and Krebs~\cite{grachten14:_asses_learn_score_featur_model}, and  Van Herwaarden et al.~\cite{vanHerwaarden:2014uc} present an alternative, unsupervised approach to modeling musical dynamics using restricted Boltzmann machines.
This approach uses a piano roll representation of musical scores to explain the musical dynamics of performed piano music.
In order to predict expressive dynamics of a score, the features learned by this model are trained in a supervised fashion  using \emph{least squares} regression.
The choice of a note-centered representation of a musical score makes this system able to model harmonic context based on relative pitch, but insensitive to absolute pitch.
Furthermore, this encoding of a score does not include performance directives written by the composer, such as dynamics or articulation markings (such as \textit{piano}, staccato, etc).
Both the KTH system and prior work on basis-function modeling have shown that the encoding of pitch and dynamics/articulation markings plays an important role in the rendering of expressive performances.

To date, there are (to the best of our knowledge) no computational models of ensemble performance in the sense that we described in Section~{\ref{sec:introduction}} above.
A slightly related method is described by \cite{xia15:_spect_learn_expres_inter_ensem_music_perfor}.
They train a model on piano duets, with the aim of predictive modeling of musical expression in order to perform automatic musical accompaniment of a human performer.

\section{A computational expression model for ensemble performance}\label{sec:comp-expr-model}
In this Section, we describe the core contribution of this paper, a computational model that predicts the intensity of a recorded ensemble performance over time, as a function of the musical score.
We begin by introducing the basis-function modeling approach (Section~\ref{sec:basis-functions}), that has been used before in a solo instrument setting.
Next, we present three variants of the basis-modeling approach: a simple linear model, and two non-linear, neural network models (Section~\ref{sec:linear-vs.-non}).
Finally, we discuss how the ensemble setting is different from the solo instrument setting, and how the basis-function modeling approach is extended to deal with ensemble performances (Section~\ref{sec:from-solo-instrument}).

\subsection{Basis-function representations of musical information}\label{sec:basis-functions}
In this section, we describe the basis-function modeling (BM) approach described by \cite{grachten12:_linear}.
In this approach, a \emph{musical score} is regarded as a set of elements on a time axis.
This set includes note elements (with attributes like pitch, duration, metrical position) as well as non-note elements (e.g.\ dynamics and articulation markings).
The set of all note elements in a score is denoted by $\mathcal{X}$.
Musical scores can be described in terms of \emph{basis-functions}, i.e.\ numeric descriptors that represent aspects of the score.
Formally, we can define a basis-function $\varphi$ as a real valued mapping $\varphi \colon \mathcal{X} \mapsto \mathbb{R}$.
In a similar way, musical expression is characterized in a quantitative way by a number of \emph{expressive parameters}.
In particular, expressive dynamics can be conveyed by the MIDI velocities of notes performed on an appropriate device such as an electronic piano or a piano equipped with sensors.
Further expressive parameters capture aspects of note timing and local tempo (e.g.
inter-onset intervals between consecutive notes), and articulation (the proportion of the duration of a note with respect to its inter-onset interval).
Although the basis-function approach can be applied without any alteration to model all of these expressive parameters, the focus in this study will be on expressive dynamics.

By defining basis-functions as functions of notes, instead of functions of time, the BM framework allows for modeling forms of music expression related to simultaneity of musical events, like the micro-timing deviations of note onsets in a chord, or the melody lead~\cite{goebl01:_melod} in piano performance, i.e.\ the accentuation of the melody voice with respect to the accompanying voices by playing it louder and slightly earlier.
However, expressive information for individual notes is difficult to obtain, and in situations where this information is not available (as in the present study), we represent expressive information as a function of time, rather than a function of notes.
We return to this issue in Section~\ref{sec:meas-vers-comp}.

\begin{figure}[t]
\begin{center}
\includegraphics[width=0.5\textwidth]{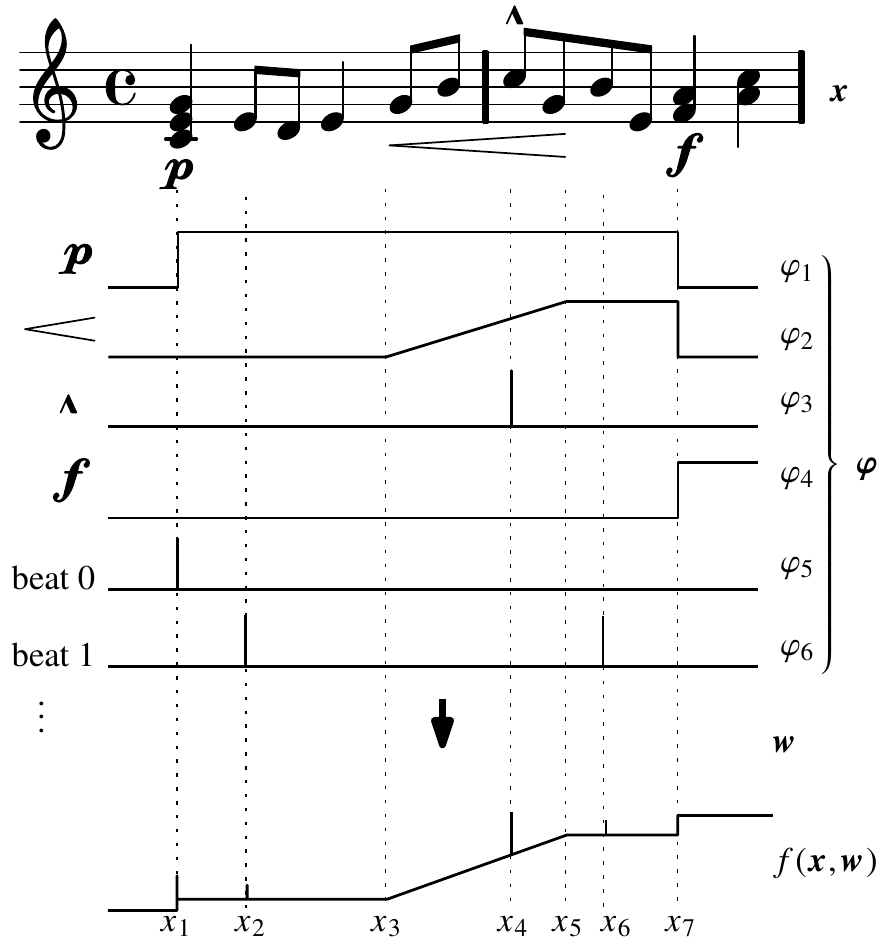}
\caption{Schematic view of expressive dynamics as a function  $f(\vex{x},\vex{w})$ of basis-functions $\varphi$, representing dynamic annotations and metrical basis functions.} \vspace{-7mm}\label{fig:basisfunctiondiagram}
\end{center}
\vskip1ex
\end{figure}

Figure~\ref{fig:basisfunctiondiagram} illustrates the idea of modeling expressive dynamics using basis-functions schematically.
Although basis-functions can be used to represent arbitrary properties of the musical score (see Section~\ref{sec:groups-basis-funct}), the BM framework was proposed with the specific aim of modeling the effect of \emph{dynamics markings}.
Such markings are hints in the musical score, to play a passage with a particular dynamical character.
For example, a \emph{p} (for \emph{piano}) tells the performer to play a particular passage softly, whereas a passage marked \emph{f} (for \emph{forte}) should be performed loudly.
Such markings, which specify a constant loudness that lasts until another such directive occurs, are modeled using a step-like function, as shown in the figure.
A gradual increase/decrease of loudness (\emph{crescendo}/\emph{diminuendo}) is indicated by right/left-oriented wedges, respectively.
Such markings are encoded by ramp-like functions.
A third class of dynamics markings, such as \emph{marcato} (i.e.\ the ``hat'' sign over a note), or textual markings like \emph{sforzato} (\emph{sfz}), or \emph{forte piano} (\emph{fp}), indicate the accentuation that note (or chord).
This class of markings is represented through (translated) unit impulse functions.
In the BM approach, the expressive dynamics are modeled as a combination of the basis-functions, as displayed in the figure.

\subsubsection{Groups of basis-functions}\label{sec:groups-basis-funct}

As stated above, the BM approach encodes a musical score into a set of numeric descriptors.
In the following, we describe various groups of basis-functions, each group representing a different aspect of the score.
This list should by no means be taken as an exhaustive (or accurate) set of features for modeling musical expression.
It is a tentative list that encodes basic information, either directly available, or easily computable from a symbolic representation of the musical piece (such as MusicXML\footnote{\url{http://www.musicxml.com}}).

\begin{enumerate}
\item \textbf{Dynamics markings}.
Bases that encode dynamics markings, such as shown in Figure~\ref{fig:basisfunctiondiagram}.
For each of the constant loudness markings (\textit{p}, \textit{pp}, \textit{f} etc.), two additional ramp-function are included that allow for a gradual change towards the loudness level indicated by the marking.
Such bases are referred to as \emph{anticipation} functions, and we distinguish between \emph{long} and \emph{short} anticipations, according to how gradual is the change towards the target dynamics marking.
Additionally, basis-functions that describe gradual changes in loudness, such as \textit{crescendo} and \textit{diminuendo}, are represented through a combination of a ramp function, followed by a constant (step) function, that continues until a new constant dynamics marking (e.g.\ \textit{f}) appears, as illustrated by $\varphi_2$ in Figure~\ref{fig:basisfunctiondiagram}.
\item \textbf{Polynomial pitch model}.
Grachten and Widmer~\cite{grachten12:_linear} proposed a third order polynomial model to describe the dependency of dynamics on pitch.
This model can be integrated in the BM approach by defining each term in the polynomial as a separate basis-function, i.e.
``pitch'', ``pitch$^2$'', and ``pitch$^3$''.
For transposing instruments, such as some of the wind instrument found in orchestras, the actual sounding pitch (concert pitch) is used.
\item \textbf{Vertical neighbors}.
Three basis-functions that evaluate to the number of simultaneous notes with lower pitches, higher pitches, or the sum of both, respectively.
\item \textbf{Duration}.
A basis-function that encodes the duration of a note.
\item \textbf{Metrical}.
Representation of the time signature of a piece, and the position of each note in the bar.
For example, the basis-function labeled \emph{4/4 beat 0} evaluates to 1 for all notes that start on the first beat in a 4/4 time signature, and to 0 otherwise. This is illustriated by $\varphi_5$ and $\varphi_6$ in Figure~\ref{fig:basisfunctiondiagram} for the first and second beat in each bar.
\item \textbf{Accent}.
Accents of individual notes or chords, such as the \emph{marcato} in Figure~\ref{fig:basisfunctiondiagram}.
\item \textbf{Staccato}.
Encodes \emph{staccato} markings on a note, an articulation indicating that a note should be temporally isolated from its successor, by shortening its duration
\end{enumerate}

\subsection{Linear vs. non-linear modeling}\label{sec:linear-vs.-non}
Basis-function modeling provides a way of representing diverse aspects of score information in a uniform way.
The next question is how this information is used to model expressive parameters.
Initial versions of the basis-function expression model used a linear model \cite{grachten11:smc}.
In a linear model, the expressive parameters are simply a weighted sum of the basis-functions, where the parameters of the model are the weights for each basis-function, to be estimated based on training data.
A major advantage of a linear model is that the link between the basis-functions and the predictions is very clear: the weight for a basis-function expresses how strong the basis-function influences the output.
This makes it easy to perform a qualitative analysis of what the model has learned, and by fitting the model on a particular piece, or on several pieces by the same performer, the weights may also capture characteristics of the expressive quality of a piece, or a performer.
See \cite{grachten11:_method} for an example of this.

The simplicity of linear modeling is at the same time a drawback.
There are two main limitations to the linear approach.
Firstly, the shape of a basis-function can only be used literally (apart from scaling and vertical translation) to approximate an expressive parameter.
For example, a \emph{crescendo} annotation is schematically represented as a \emph{ramp function}, and this means that any increase of loudness in that region can only be approximated as a linear slope.
In reality, it is likely that the shape of the loudness increase is not strictly linear.
Secondly, the linear approach does not model any interactions between basis-functions.

To overcome these limitations, Cancino and Grachten~\cite{chacon15} proposed a non-linear basis-function model for expression, based on a \emph{feed-forward neural network}, where they ran experiments on Chopin piano music.
The Discussion Section of that paper provides an example that shows the benefit of the non-linearity of the model, both in the non-linear transformation of the basis-functions, and in the interaction between basis-functions.
More specifically, the example shows that the non-linear model reduces the effect of a crescendo in situations where the crescendo sign is directly preceded by a diminuendo sign.
Such interactions are not possible in a linear model.
The example also shows that the ramp shape of the crescendo is slightly smoothed.
We refer to the paper for more details.

A more powerful type of non-linear modeling can be obtained by introducing recurrence relations to the neural network architecture:
\emph{Recurrent Neural Networks} (RNNs) are a particular kind of  discrete--time dynamical artificial neural networks (ANNs) suited for analyzing sequential data, such as time-series.
These dynamic models have been successfully used for generating text sequences, handwriting synthesis and modeling motion capture data~\cite{Graves:2013sm}.
The structure of an RNN is similar to that of a feed forward neural network, with the particularity that it allows connections among its states associated with time delays.
It is through these connections that RNNs are able to capture temporal correlations between events~\cite{Pascanu:2014vu}.

\subsection{From solo piano to orchestral ensembles}\label{sec:from-solo-instrument}
The basis-function modeling approach described above has been developed for the purpose of modeling expression in solo piano performances, based on precise measurements obtained from a computer-controlled grand piano.
There are several issues to be dealt with in order to apply the same approach to orchestral performances.
In the rest of this Section, we will discuss these issues, and provide solutions.

\subsubsection{Measured versus computed expressive parameters}\label{sec:meas-vers-comp}
In a piano, the degrees of freedom for sound production, and therefore expressive performance, are limited to only a few, well-defined dimensions (such as hammer velocity, timing of key press and release)
that can be measured relatively easily.
Through the use of computer-controlled pianos \cite{moog90}, it is possible to obtain precise measurements of these dimensions in piano performances.
Similar measurements are typically not easily possible for other classes of instruments, such as bowed string instruments, or wind instruments, which have more complex sound production mechanisms.
Although with the appropriate sensors, rich descriptions of non-piano performances may be obtained (for instance to measure the bending of the reed in wind instruments \cite{hofmann2013measurement} or bow movements in violin playing~\cite{schoonderwaldt_2009_bowing_parameters}),
the usage of such sensors is often intrusive, and thus limited to experimental setups.
Moreover, data recorded in this way is prone to noise, and bulky in case of large ensembles.

For these reasons, our current work is focused on relatively coarse, but easy to obtain form of expressive information, namely the instantaneous overall loudness computed from an audio recording of a professional music performance.
This implies that there is only a single value for each expressive parameter at each \emph{time instant}, as opposed to the measured piano scenario, where expressive parameters can be defined in part for \emph{individual notes}, even if they occur at the same time instant. Since the basis-functions of the form described in Section~\ref{sec:basis-functions} return a value for each note, and thus possibly multiple values for a single time instant, it is necessary to fuse these values in order to obtain a single prediction for the expressive parameter at that time instant.

\subsubsection{Indexing basis-functions}\label{sec:index-basis-funct-1}
The basis-modeling approach, including the list of basis-functions defined in Section~{\ref{sec:basis-functions}}, is designed to generate a set of basis-functions, given a \emph{score part} for an instrument.
When training an expression model on a data set containing performances of multiple pieces, the basis-functions produced for each piece must be mapped to each other.
In the solo instrument setting, this mapping is done on the basis of labels that are uniquely assigned to each basis-function.
In orchestral scores, the labels are not unique any longer, since the same basis-functions are produced for each instrument (coding the same type of score information, but for different instruments).
To deal with this, it is necessary to index the basis-functions by the tuple \textit{(instrument name, basis-function label)}.

A further issue is that the notated instrument names in the score, do not follow any strict standard.
Instrument names may be written in different languages (e.g.
``Fagott'', ``bassoon''), and may be abbreviated (e.g.
``Vln.'' for violin, ``Cl.'' for clarinet).
To overcome this issue, the instrument names and abbreviations extracted from the score are matched to one of a set of \emph{canonical instrument names} (the unabbreviated English names), using string matching techniques.

\subsubsection{Merging and fusion of basis-functions within instrument classes}\label{sec:fusi-basis-funct}

In orchestral scores, there may be several instances (voices) of an instrument, usually designated by numbers (e.g.
``Violin 1'', ``Violin 2'').
Furthermore, multiple instances of an instrument may share a single staff.

The occurrence of multiple instruments of the same type poses a problem for training the model, since it is not clear how the mapping of basis-functions across pieces should be defined in order to create a consistent dataset consisting of multiple pieces.
For instance, when one piece involves a single violin, and another piece involves two violins, the question which of the two violins in the latter piece should be mapped to the violin in the first piece is arbitrary, and moreover, it is unclear how to deal with the remaining, unmapped violin.

For this reason, we choose to combine all instances of the same instrument class into a single set of basis-functions, using a \emph{fusion operation} that can be specified per basis-function type.
In this way, for each piece there is a single set of basis-functions conveying the activity of a given instrument \emph{class}, rather than one set for each \emph{instance} of that class.

\begin{figure*}[htb]
  \centering
  \includegraphics[width=0.8\linewidth]{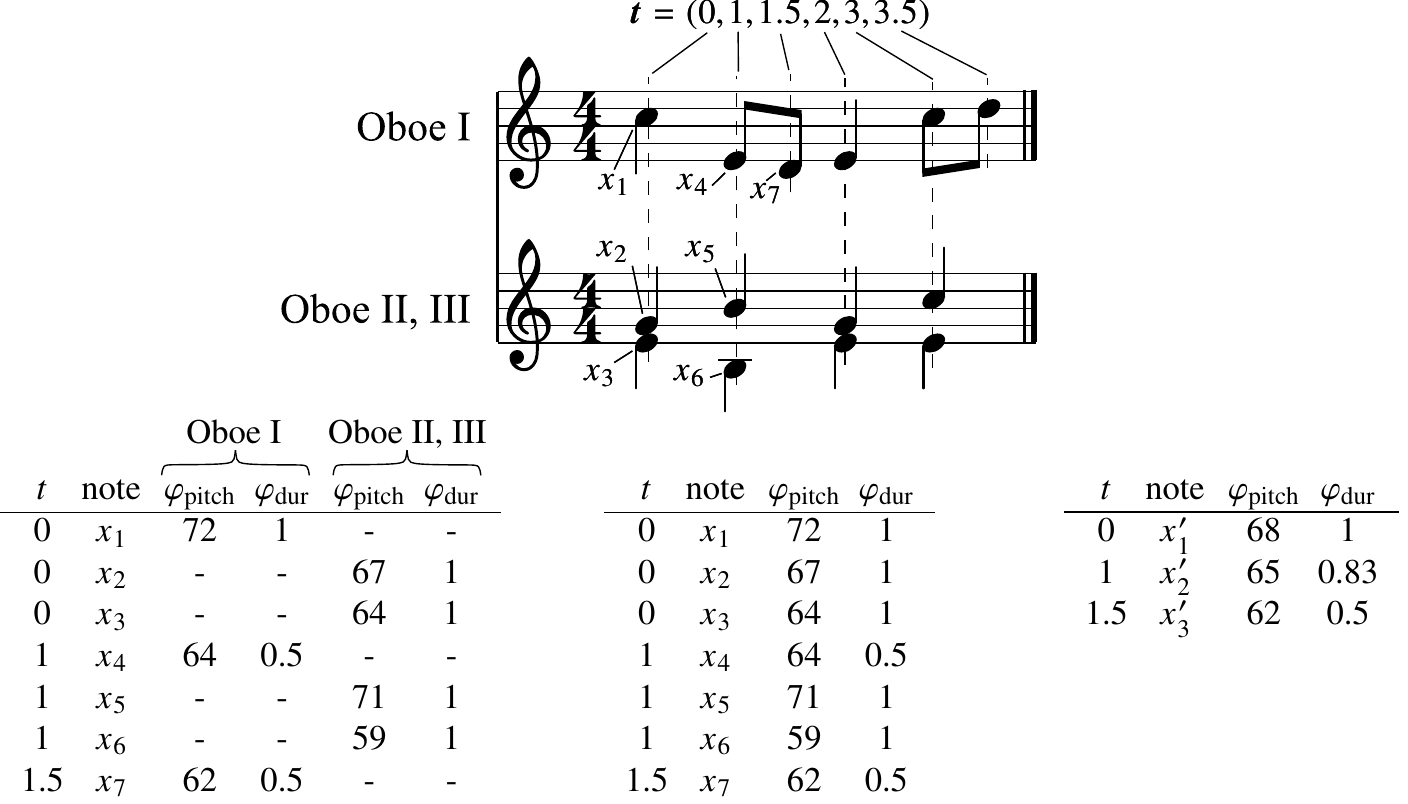}
  \caption{Illustration of \emph{merging} and \emph{fusion} of score information of two different parts belonging to the same instrument class \enquote{Oboe}.
  The matrix on the left shows the basis-functions $\varphi_{\mathrm{pitch}}$ and $\varphi_{\mathrm{dur}}$ for each of the two score parts (truncated after the first few notes).
  Note the consecutive layout of simultaneously occurring notes, as indicated by the first column of each matrix, giving the notes' onset times.
  The matrix in the center illustrates \emph{merging}, where the data of both score parts were combined.
  Finally, the third matrix on the right is the result of \emph{fusion} operations, applied per basis function to each set of values occurring at the same time point.
  See text for further explanation.}
  \label{fig:merge_and_fuse}
\end{figure*}

The process of merging and fusion is shown in Figure~\ref{fig:merge_and_fuse}.
First, a collection of $K$ predefined basis-functions $\vex{\varPhi} = (\vex{\varphi}_1, \vex{\varphi}_2, \ldots, \vex{\varphi}_K)$ is applied to each score part, where e.g. $\varphi_1 := \varphi_{\mathrm{pitch}}$ corresponds to the pitch of a note, expressed as the MIDI note number.
This gives a matrix of basis-function values for each score part.
Notes occurring at the same time are laid out consecutively, as shown in the leftmost matrix in Figure~\ref{fig:merge_and_fuse};
Note that the two score parts' matrices were already stacked together here, with the rows arranged according to the notes' onset times.

Second, the score information of different score parts belonging to the same instrument class, here \enquote{Oboe}, need to be combined into a single instrument class matrix.
This is referred to as \emph{merging}.
As can be seen from the matrix in the center, the corresponding columns of the leftmost matrix were stacked into one column each, with simultaneous notes still consecutively listed.
The number of columns is the cardinality of the set of the basis-functions of the involved score parts.

Finally, a fusion operation is applied to each subset of a column having the same onset time $t$, yielding a single value $\varphi$ for each time instant.
The matrix on the right in Figure~\ref{fig:merge_and_fuse} results from applying fusion to the matrix in the center.
The number of rows is given by the size of the union of all occurring onset times.
Following this procedure, for each instrument class in a piece, there will be a single collection of basis-functions that can easily be mapped to other pieces' basis function of the same instrument class.
Thereby, a collection of matrices $\vex{\varPhi}_i$ for the instrument classes $i = 1, 2, \ldots, I$ of a piece is produced.

\subsubsection{Aggregation of basis-functions of instrument classes in a piece}\label{sec:aggreg-basis-funct}

After collecting per instrument class basis-function matrices, we need one final step to conclude the data extraction for a piece $\mathcal{P}$.
All instrument classes' data are aggregated into a single per-piece matrix $\vex{\varPhi}_{\mathcal{P}}$.
The number of rows of this matrix is given by the total number of unique onset times across all score parts $P$ and is denoted $N_{\mathcal{P}}$.
The number of columns of $\vex{\varPhi}_{\mathcal{P}}$ is given by $K_{\mathcal{P}}$, the sum of the number of columns of the single instrument class matrices $\vex{\varPhi}_i$, thus $\vex{\varPhi}_{\mathcal{P}} \in \mathbb{R}^{N_{\mathcal{P}} \times K_{\mathcal{P}}}$.

\subsubsection{Model description}\label{sec:model-description}

For the training procedure across multiple pieces, it is necessary to match all the pieces' basis-functions to each other.
This can again be achieved by appropriately stacking together all involved $\vex{\varPhi}_{\mathcal{P}}$ to produce a data set matrix $\vex{\varPhi}_{\mathcal{S}}$ of shape ($N_{\mathcal{S}} \times K_{\mathcal{S}}$). $N_{\mathcal{S}}$ is the sum of the number of rows of all per-piece matrices $\vex{\varPhi}_{\mathcal{P}}$, whereas $K_{\mathcal{S}}$ is the cardinality of the set of all uniquely occurring basis functions across the data set.

The model can now be described in the following way.
In general, an expressive target parameter $\vex{y}$ is modeled as a function $f(\cdot)$ of the data $\vex{\varPhi}_{\mathcal{S}}$ extracted from the scores and a vector $\vex{w}$ of weights:

\begin{equation}
\vex{y} = f(\vex{\varPhi}_{\mathcal{S}}, \, \vex{w}) + \vex{\varepsilon}
\end{equation}
Here, $\vex{w}$ has shape ($K_{\mathcal{S}} \times 1$), and $\vex{\varepsilon}$ is zero mean Gaussian noise with covariance matrix $\sigma^2 \vex{I}$, with $\vex{I}$ an identity matrix of appropriate shape.

In a linear model, the function $f(\cdot)$ is a linear combination:
\begin{equation}
\vex{y} = \vex{\varPhi}_{\mathcal{S}} \, \vex{w} + \vex{\varepsilon}.
\end{equation}

In the following Section \ref{sec:experiments}, we describe the experiments that used the linear model, a Feed Forward Neural Network and a Recurrent Neural Network (RNN) for estimating the model parameters (weights) $\vex{\hat{w}}$.
Once the estimated parameters are established, they can be used to predict the loudness variations $\vex{\hat{y}}_{\mathcal{P}}$ from the score information $\vex{\varPhi}_{\mathcal{P}}$ of a piece (not yet seen by the model):
\begin{equation}
\vex{\hat{y}}_{\mathcal{P}} = f(\vex{\varPhi}_{\mathcal{P}}, \vex{\hat{w}}).
\end{equation}

We will not go into more details about the model here but instead refer the interested reader to Cancino and Grachten \cite{chacon15} for a more detailed and formal explanation.

\section{Experiments}\label{sec:experiments}

In our experiment, we want to assess how well we can predict variations in loudness (as an expressive parameter) of ensemble pieces using the basis-function model.
We compare different models, namely a linear model, a Feed Forward Neural Network (FFNN) and a Recurrent Neural Network (RNN).

\subsection{Data}\label{sec:data}
The corpus used for the experiments is summarized in Table \ref{table:data} below. It consists of symphonies from the classic and romantic period.
\begin{table}[h]
\footnotesize
\centering
\caption{Pieces used in the experiments.}
\begin{tabular}{lllll}
\toprule
Composer & Piece & Cond. & Movements \\
\midrule
Beethoven & S. 6 in F-Maj. (op. 68)  & Fischer & 1, 2, 3, 4, 5 \\
Beethoven & S. 9 in D-Min. (op. 125) & Fischer & 1, 2, 3, 4 \\
Mahler    & S. 4 in G-Maj.           & Jansons & 1, 2, 3, 4 \\
Bruckner  & S. 9 in D-Min. (WAB 109) & Jansons & 1, 2, 3 \\
\bottomrule
\end{tabular}
\label{table:data}
\end{table}

For each of these symphonies, a recorded performance (an audio file), a machine-readable representation of the musical score (a MusicXML file) and an automatically produced, manually corrected alignment between score and performance are available in the corpus.

We used recordings of performances by the Royal Concertgebouw Orchestra conducted by Ivan Fischer or Mariss Jansons, all performed at the Royal Concertgebouw in Amsterdam, the Netherlands.
Since the various movements of a symphony are handled individually, from now on each movement is referred to as a piece.
The corpus thus amounts to a total of 16 pieces.
The corresponding performances sum up to a total length of almost 4 hours of music.
From the 16 pieces' scores, a total of $N_{\mathcal{S}} = 47228$ note onsets, belonging to $K_{\mathcal{S}} = 1518$ basis functions, were extracted.

The symbolic scores used for the extraction of the basis-functions were provided partly by B{\"a}renreiter Verlag\footnote{\url{http://www.baerenreiter.com}}, and partly by Donemus Publishing\footnote{\url{http://www.donemus.nl}}.
The target values (loudness) for each piece were extracted using the loudness measure defined by EBU R128, as described in~\cite{ebur128}.

The recordings were all made in the same hall and produced for the same target medium. Thus, we do not expect the recording and production process to be a significant source of variation in loudness.

To map the note onset times in the score--the positions at which basis functions are evaluated--to loudness values in the recorded performance,
the score-to-audio alignment technology as described in \cite{grachten_ismir_2013} was used.
The alignments were corrected by a human annotator at least at the level of single bars.
It makes sense to estimate the loudness corresponding to a particular score note by measuring the loudness slightly \emph{after} the estimated onset time in the performance.
One reason for this is that some instruments have a significant attack-time, meaning that peak loudness occurs some time after the start of the note.
Another reason is that in the possible case of some minor residual error in the alignment after correction, the probability that a particular note estimated to start at $t$ is actually sounding is higher at $t + \delta$ than at $t$ for some positive $\delta$, assuming sum of the average residual alignment error and the chosen $\delta$ is smaller than the average note duration.
For these reasons, we extracted the target value 1/10th of a beat after the onset time point given by the alignment to decrease the probability of \enquote{hitting} a note before its onset.

\subsection{Method}\label{sec:method-1}

We used a leave-one-out scenario where the model is trained on 15 of the 16 pieces and then is used to predict the target values for the unseen remaining piece.
The non-linear models (FFNN, and RNN) are trained by gradient descent optimization.
Both the feed-forward and the recurrent neural network each were set up with a single hidden layer of 20 units.
From the 15 training pieces, two pieces were kept for validation, to avoid overfitting the models to the training data, a practice known as \emph{early stopping}~\cite{NIPS1989_275}.
The predictions are evaluated with respect to the target (here loudness curve) in terms of the Pearson correlation coefficient $r$ and the coefficient of determination $R^2$.

The set of basis-functions used in the experiments encode note pitch, duration, and metrical position, the number of simultaneous notes within instrument groups, inter-onset intervals between subsequent notes, repeat signs, note accent, staccato, fermata signs, and dynamics markings. A full description of the basis functions is omitted for brevity.

\subsection{Results and discussion}\label{sec:results-discussion}
The results of the experiments are shown in the following Table~\ref{table:results}.
For each piece, we report the \emph{Coefficient of determination} ($R^2$) that measures the proportion of variance in the recorded loudness curve that is explained by the model,
and \emph{Pearson's correlation coefficient} ($r$), that measures the strength of the linear dependence between the recorded and the predicted loudness curves.
The $R^2$ measure has an upper bound of 1, and has no lower bound (predictions can be arbitrarily far away from the target values).
Positive $R^2$ values indicate that the models perform better than the baseline of predicting the mean value of the loudness over the whole piece.

\begin{table*}[]
\footnotesize
\caption{Predictive accuracy in a leave-one-out scenario for different models;
MSE = mean squared error (smaller is better); $R^{2}$ = coefficient of determination (larger is better); $r$ = Pearson correlation coefficient (larger is better);
RN = recurrent neural network; FF = feed forward neural network; Lin = linear model;
Best value per piece and measure emphasized in bold}
\label{table:results}
\newcommand{\inw}{\hskip1em}
\newcommand{\outw}{\hskip2.8em}
\newcommand{\thead}[2]{\multicolumn{1}{c@{#1}}{#2}}
\begin{center}
\begin{tabular}{
l@{\hskip.4em}l@{\outw}
r@{\inw}r@{\inw}r@{\outw}
r@{\inw}r@{\inw}r@{\outw}
r@{\inw}r@{\inw}r}
\toprule
Piece & & 
\multicolumn{3}{c@{\outw}}{MSE} & 
\multicolumn{3}{c@{\outw}}{$R^2$} & 
\multicolumn{3}{c}{$r$} \\ 

      & & 
\thead{\inw}{RN} & \thead{\inw}{FF} & \thead{\outw}{Lin} & 
\thead{\inw}{RN} & \thead{\inw}{FF} & \thead{\outw}{Lin} & 
\thead{\inw}{RN} & \thead{\inw}{FF} & \thead{\inw}{Lin} \\ 
\midrule

LvB S6      & Mv 1 & \textbf{0.57} & 0.60 & 0.96  & \textbf{0.43} & 0.40 & 0.04   & \textbf{0.71} & 0.66          & 0.40 \\
            & Mv 2 & \textbf{0.80} & 0.87 & 0.94  & \textbf{0.20} & 0.13 & 0.06   & \textbf{0.45} & 0.36          & 0.35 \\
            & Mv 3 & \textbf{0.40} & 0.45 & 0.56  & \textbf{0.60} & 0.55 & 0.44   & \textbf{0.79} & 0.76          & 0.67 \\
            & Mv 4 & \textbf{0.66} & 0.67 & 0.87  & \textbf{0.34} & 0.33 & 0.13   & \textbf{0.61} & 0.60          & 0.42 \\
            & Mv 5 & \textbf{0.52} & 0.59 & 0.66  & \textbf{0.48} & 0.41 & 0.34   & \textbf{0.74} & 0.68          & 0.58 \\
Mah S4      & Mv 1 & \textbf{0.64} & 0.76 & 6.21  & \textbf{0.36} & 0.24 & -5.21  & \textbf{0.60} & 0.51          & 0.02 \\
            & Mv 2 & \textbf{0.95} & 0.98 & 11.69 & \textbf{0.05} & 0.02 & -10.69 & \textbf{0.26} & 0.22          & 0.03 \\
            & Mv 3 & \textbf{0.51} & 0.66 & 2.63  & \textbf{0.49} & 0.34 & -1.63  & \textbf{0.71} & 0.59          & 0.19 \\
            & Mv 4 & \textbf{0.86} & 0.96 & 2.03  & \textbf{0.14} & 0.04 & -1.03  & \textbf{0.40} & 0.29          & 0.18 \\
LvB S9      & Mv 1 & \textbf{0.62} & 0.67 & 0.70  & \textbf{0.38} & 0.33 & 0.30   & \textbf{0.63} & 0.60          & 0.58 \\
            & Mv 2 & \textbf{0.44} & 0.57 & 0.67  & \textbf{0.56} & 0.43 & 0.33   & \textbf{0.75} & 0.66          & 0.59 \\
            & Mv 3 & \textbf{0.95} & 0.98 & 1.24  & \textbf{0.05} & 0.02 & -0.24  & 0.27          & \textbf{0.29} & 0.26 \\
            & Mv 4 & \textbf{0.62} & 0.80 & 0.92  & \textbf{0.38} & 0.20 & 0.08   & \textbf{0.63} & 0.55          & 0.45 \\
Bru S9      & Mv 1 & \textbf{0.41} & 0.52 & 7.22  & \textbf{0.59} & 0.48 & -6.22  & \textbf{0.77} & 0.70          & 0.24 \\
            & Mv 2 & \textbf{0.39} & 0.48 & 0.97  & \textbf{0.61} & 0.52 & 0.03   & \textbf{0.80} & 0.74          & 0.47 \\
            & Mv 3 & \textbf{0.61} & 0.65 & 0.99  & \textbf{0.39} & 0.35 & 0.01   & \textbf{0.65} & 0.59          & 0.33 \\

\bottomrule
\end{tabular}
\end{center}
\end{table*}

Several observations can be made from Table~\ref{table:results}.
First of all, both the $R^2$ and the $r$ values for Lin are generally lower than those for FFNN and RNN, demonstrating that the non-linear modeling provides a clear advantage over the linear modeling approach.
Given the relatively small data set, this is surprising, since the FFNN and RNN have much more parameters than the Lin model, and are therefore more prone to both \emph{overfitting} or \emph{underfitting}.
Secondly, the RNN model provides more accurate predictions than the FFNN model, although this advantage is less prominent than the advantage over the linear model.

A possible explanation for the advantage of the RNN model lies in a limitation of the basis-function modeling approach in its current form.
The example in Figure~\ref{fig:merge_and_fuse} illustrates this limitation. Notes $x_5$ and $x_6$, starting at beat 1 are quarter notes, still sounding at the onset of note $x_7$ at beat 1.5.
Thus, it is to be expected that the presence of notes $x_5$ and $x_6$ will affect the overall loudness value at beat 1.5.
However, the basis-functions representing those notes are only active at beat 1, and not at beat 1.5 (the only row pertaining to beat 1.5 in the left matrix is the one representing $x_7$).
The linear model and the FFNN have no way to incorporate basis-function information describing notes $x_5$ and $x_6$ at time 1.5, but through its recurrent connections, the RNN \emph{can} learn that information at prior time steps can be helpful to predict the loudness at the current time step.
To verify this explanation, a further analysis of the results is necessary, which is beyond the scope of this paper.

Furthermore, Table~\ref{table:results} shows that all models have difficulty predicting the loudness curves for some of the pieces, in particular for the 2nd movement of Mahler's 4th Symphony and the 3rd movement of Beethoven's Symphony No. 9.
Since the data set is relatively small, we hypothesized that the inaccurate predictions for these pieces might be due to the occurrence of \emph{singular} basis-functions: basis-functions that are active in the test piece, but not (or hardly) active in any training piece.
This may result in an undertraining of the models for these basis-functions, leading to inaccurate predictions of the loudness curves.
However, upon testing this, we found that the pieces with low predictive accuracy did not have substantially larger numbers of singular basis-functions than other pieces.
For further investigation a sensitivity analysis may be helpful, to test whether the models are more sensitive to singular basis-functions for the problematic pieces than for other pieces.
Should this be the case, the models may benefit from stronger regularization of the model parameters during training.

\begin{figure*}[Htb]
\centering
\includegraphics[width=.98\linewidth]{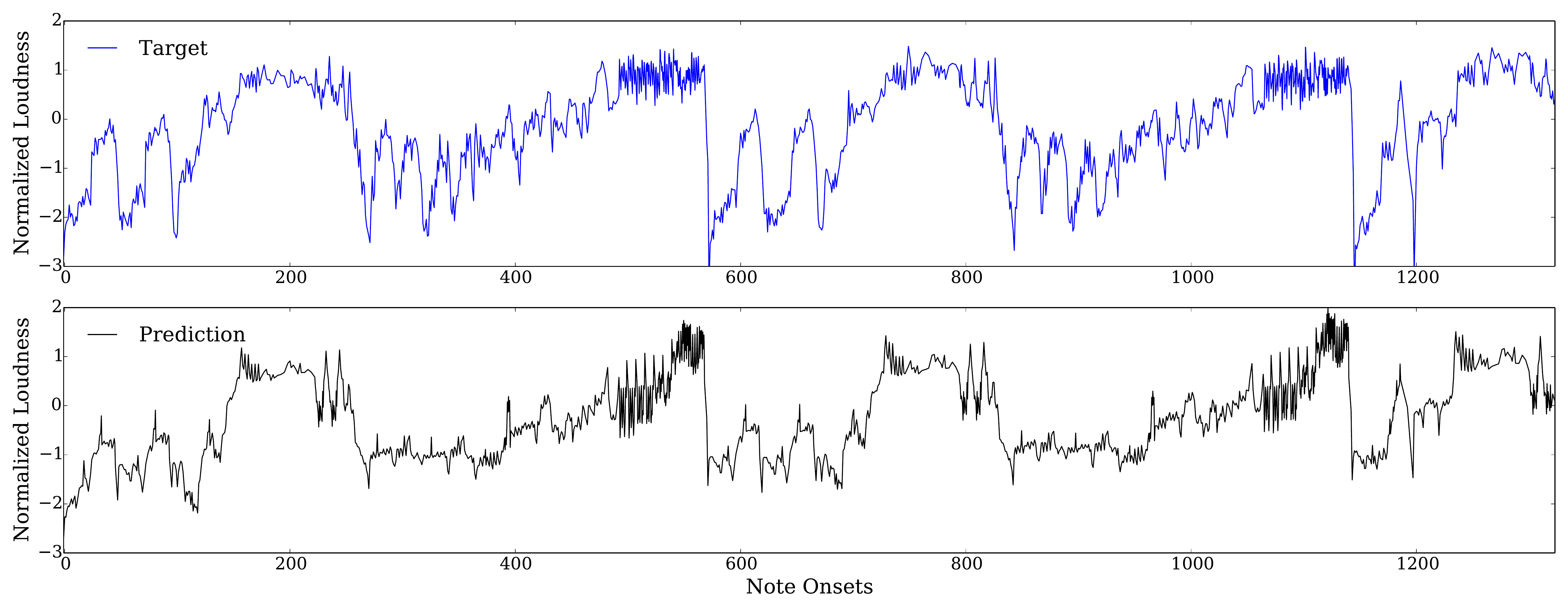}
\caption{Prediction of loudness variation in movement 3 of Beethoven's Symphony No. 6.
The upper curve shows the (normalized) loudness extracted from the audio recording, the lower curve the loudness variations predicted by the recurrent neural network (RNN), based on the written score.
}
\label{fig:prediction_new}
\end{figure*}

Figure \ref{fig:prediction_new} shows an example of a loudness curve extracted from a recorded performance, and the predicted loudness curve by the RNN (trained on other pieces), based on the written score.
Note that although the details of the predicted loudness curve are not very accurate, the overall shape of the predicted curve resembles the actual loudness curve.

\section{Conclusion and future work}\label{sec:conclusion}
In this paper, we have described an extension of an existing model for musical expression for solo piano to deal with performances of ensembles, such as a symphonic orchestra.
The model represents score information for each instrument in the ensemble, and uses the joint information to predict the overall loudness curve of a recorded performance.
We have evaluated three variants of the model (one linear version and two non-linear versions), on a dataset of recorded performances of symphonic music pieces by Mahler, Beethoven, and Bruckner, played by the Royal Concertgebouw Orchestra.
Although the data set is rather small, the experiments show that the non-linear models have a clear advantage over the linear model, and also that the recurrent non-linear model performs better than the feed forward non-linear model.

Arguably the overall loudness curve of a performance is a very coarse way of representing expressive variation of dynamics.
For more precise and reliable modeling, it is desirable to have loudness values available per instrument, or per instrument class.
A set of possibly useful recordings (where each instrument of the orchestra is recorded in isolation) is reported in \cite{paetynen_2008}.
For pragmatic and technical reasons however, it is unfeasible to record each instrument separately in live performances.

Source separation techniques such as reported in \cite{marxer2012low}, and \cite{miron2015improving}, may also be useful for more precise expression modeling, as they provide a means to separate instrument families from an orchestral recording, to be used for modeling loudness of individual instrument sections.

\acknowledgements{
This work is supported by the European Union's Seventh Framework Programme FP7 / 2007-2013 (projects PHENICX / grant number 601166 and Lrn2Cre8 / grant number 610859), and by the European Research Council (ERC) under the EU's Horizon 2020 Framework Programme (ERC Grant Agreement number 670035, project CON ESPRESSIONE).
Furthermore, we wish to thank the Royal Concertgebouw Orchestra, in particular Marcel van Tilburg and David Bazen, for providing the audio recordings used in this study.}

\normalsize
\section*{\hskip1.5em References}

\begingroup
\renewcommand{\section}[2]{}
\bibliographystyle{unsrt}

\IfFileExists{bibliographies/bib_mg.bib}{
\bibliography{bibliographies/bib_mg,bibliographies/bib_cc,bib_tg}
}{
\bibliography{bib_mg,bib_tg}
}

\endgroup

\end{document}